# Blocker effect on diffusion resistance of a membrane channel. Dependence on the blocker geometry

Leonardo Dagdug,[1,2] Alexei T. Skvortsov,[3] Alexander M. Berezhkovskii,[2] and Sergey M. Bezrukov[2]

[1]*Departamento de Fisica, Universidad Autonoma Metropolitana-Iztapalapa, 09340 Mexico City, Mexico*

[2] *Section of Molecular Transport, Eunice Kennedy Shriver National Institute of Child Health and Human Development, National Institutes of Health, Bethesda, Maryland 20819, USA*

[3]*Defence Science and Technology Group, Melbourne, VIC 3207, Australia*

**Abstract**

Being motivated by recent progress in nanopore sensing, we develop a theory of the effect of large analytes, or blockers, trapped within the nanopore confines, on diffusion flow of small solutes. The focus is on the nanopore diffusion resistance which is the ratio of the solute concentration difference in the reservoirs connected by the nanopore to the solute flux driven by this difference. Analytical expressions for the diffusion resistance are derived for a cylindrically symmetric blocker whose axis coincides with the axis of a cylindrical nanopore in two limiting cases where the blocker radius changes either smoothly or abruptly. Comparison of our theoretical predictions with the results obtained from Brownian dynamics simulations shows good agreement between the two.

# 1. Introduction

Partial time-resolved blockage of ion currents through nanopores is a source of information in the rapidly developing field of nanopore-based sensing. It is emerging as a powerful tool for detection and analysis of analytes of diverse origin at a single molecule level. A substantial progress has already been achieved in DNA and protein sequencing,[1] investigations of protein-protein interactions,[2] and studies in single-molecule enzymology and post-translational protein modifications.[3] Nanopore sensing is based on the fact that upon their passage through a nanopore different analyte molecules reduce, that is, partially block, the nanopore current to a different extent, thus allowing to gauge their physico-chemical properties. Though there is a well-appreciated progress in quantitative understanding of the amplitude and time characteristics of the analyte-induced transients in nanopore current[4], a reliable theoretical background of this phenomenon is still missing. The difficulty is that any comprehensive theory of the effect of the blocking analyte molecule on nanopore current must account for multiple factors. In addition to the size and geometry of the blocking analyte, there are charge distributions on both analyte and nanopore wall surfaces, state of the water hydrating these surfaces[5], networks of hydrogen bonding, and many others. Indeed, all these complications are less severe for the large synthetic nanopores[6] but could still be important. This leads to a situation when researchers often have to rely on empirical findings. In the present study we undertake a step to developing analytical tools for assessing the blocking effect by focusing on only one of the above-mentioned factors, namely, the geometric or steric constraints imposed by the trapped analyte molecule. More specifically, we study how the presence of a large molecule in a cylindrical nanopore affects the diffusion flow of small neutral solutes, quantified in terms of nanopore diffusion resistance. Using this concept, we recently derived an analytical result for the diffusion flux through a cylindrical nanopore containing a thin partition with a circular opening of an arbitrary radius in its center.[7] We hope that our study provides a baseline for thinking about the relation between geometrical parameters of the blocker and its effects on nanopore-facilitated transport.

As a simple model, consider solutes, represented by point particles, diffusing in two reservoirs separated by a membrane. Particles can pass from one reservoir to another through a channel (or nanopore) connecting the reservoirs. When the particle concentrations in the reservoirs are different, there is a steady-state flux $J$ flowing through the channel, which can be written as[7-8]

$$J = \frac{c_{left} - c_{right}}{R_{dif}}, \qquad (1.1)$$

where $c_{left}$ and $c_{right}$ are the particle concentrations in the left and right reservoirs, respectively, and $R_{dif}$ is the channel diffusion resistance. At high particle concentrations in the reservoirs, when interparticle interactions are starting to be involved in an essential way, $R_{dif}$ is a function of these concentrations. However, at low concentrations, $R_{dif}$ is a concentration-independent function of the channel geometry and the particle interaction with the channel. Under such conditions, as follows from Eq. (1.1), the flux $J$ is the difference of independent fluxes flowing through the channel in opposite directions. With this in mind, we focus on the left-to-right flux, assuming that $c_{right} = 0$, and hence the right-to-left flux is zero. The concentration in the left reservoir below is denoted by $c$, $c_{left} = c$.

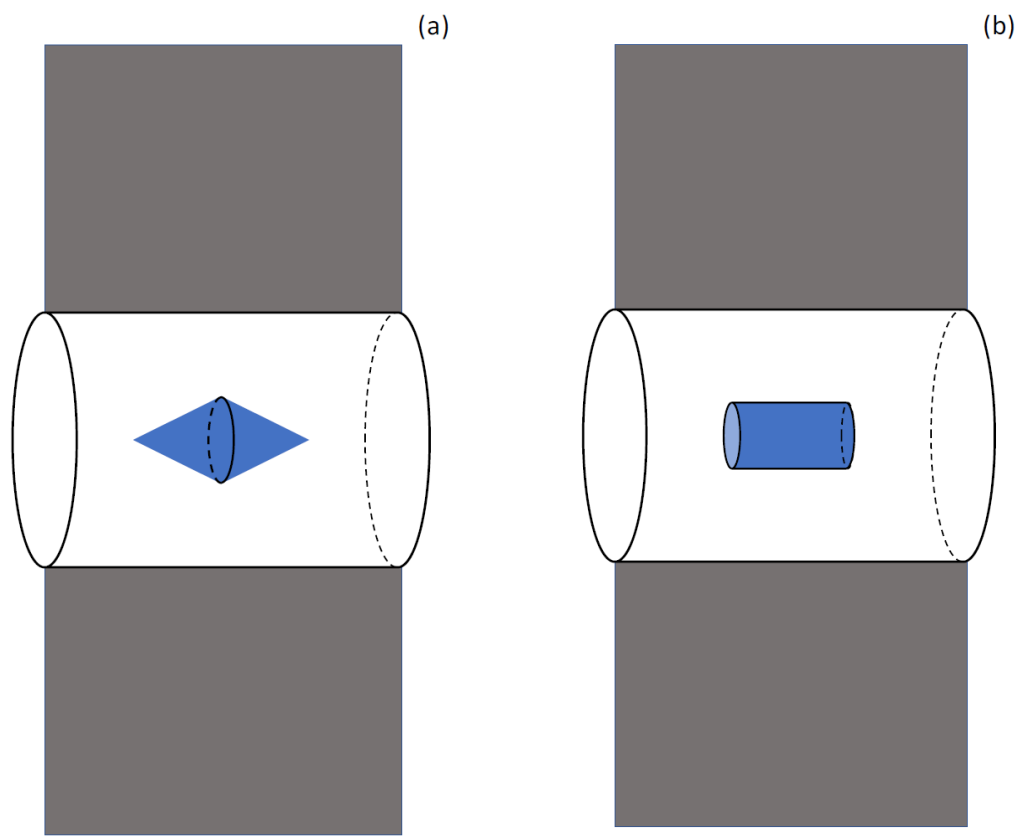

**Fig. 1 (a), (b).** Schematic representation of a cylindrical channel with a blocker formed by two cones connected by their bases (panel (a)) and with a blocking cylinder (panel (b)).

When a finite-size analyte molecule, a blocker, enters the channel, the particle flux decreases, as the blocker partially (or completely) obstructs the channel. This leads to an increase in the channel diffusion resistance. In the present study we analyze how this increase depends on blocker's shape and size in the case of a cylindrically symmetric blocker located in a cylindrical channel of constant radius (Fig. 1). Assuming that blocker's diffusivity is much lower than that of the particles, we consider the blocker as a static obstacle for the diffusing particles. As explained below, analytical expressions for $R_{dif}$ can be derived when the blocker is cylindrically symmetric, and its axis coincides with the axis of the channel. The results can be obtained in the two limiting cases, where the blocker radius changes smoothly, as exemplified by two cones connected by their bases in Fig. 1 (a), and abruptly, as represented by a blocking cylinder in Fig. 1 (b).

As an introduction of our approach to the problem consider a steady-state flux through a straight cylindrical channel of radius $R$ and length $L$ without a blocker. Let $J_{in}$ be a steady-state flux of new particles entering the channel from the left reservoir for the first time. This flux is given by

$$J_{in} = k_{HBP}c, \qquad k_{HBP} = 4RD_b, \qquad (1.2)$$

where $k_{HBP}$ is the Hill-Berg-Purcell rate constant[9] that describes trapping of diffusing particles by an absorbing circular disk located on the otherwise reflecting flat wall, and $D_b$ is the bulk diffusivity of the particles in the reservoirs. Introducing the particle translocation probability, $P_{tr}$, which is the probability that a particle entering the channel will cross it and escape to the right reservoir, we can write the steady-state flux through the channel as

$$J = J_{in} P_{tr} = 4RD_b P_{tr} c. \tag{1.3}$$

Comparison of the above equation and Eq. (1.1) with $c_{right} = 0$ and $c_{left} = c$ shows that

$$R_{dif} = \frac{1}{4RD_b P_{tr}}. \tag{1.4}$$

Thus, to find $R_{dif}$ we need to know $P_{tr}$.

The translocation probability for a cylindrical channel is given by[10]

$$P_{tr}^{(cyl)} = \frac{1}{2 + \frac{4LD_b}{\pi RD_{ch}}}, \tag{1.5}$$

where $D_{ch}$ is the particle diffusivity in the channel, which can differ from $D_b$. Substituting this into Eq. (1.4), we arrive at the expression

$$R_{dif}^{(cyl)} = \frac{1}{2RD_b} + \frac{L}{\pi R^2 D_{ch}} \tag{1.6}$$

that gives $R_{dif}^{(cyl)}$ as the sum of two terms of different physical origin. The first one, further denoted by $R_{acc}$, is the so-called access resistance,

$$R_{acc} = \frac{1}{2RD_b}. \tag{1.7}$$

This term is associated with the particle entrance in and escape from the channel. This is why it is proportional to $1/D_b$ and independent of $D_{ch}$. The channel access resistance is the sum of access resistances characterizing the two channel openings. For a cylindrical channel these two access resistances are equal to one another, and each of them is given by $1/4RD_b = 1/k_{HBP}$.[11] Thus, the access resistance is half of the diffusion resistance of the aperture of radius $R$ in an infinitely thin partition (membrane), $L = 0$.

The second term in Eq. (1.6) is the intrinsic diffusion resistance of the channel per se. This is why this term is proportional to $1/D_{ch}$ and independent of $D_b$. When $D_b \to \infty$ or when the solutions in the two reservoirs are well stirred ($ws$), there is no difference in the particle concentration near the channel ends and in the bulk of the reservoirs. As a consequence, under such conditions the channel access resistance vanishes and Eq. (1.6) reduces to

$$R_{dif}^{(cyl)}\bigg|_{D_b \to \infty} = R_{ws}^{(cyl)} = \frac{L}{\pi R^2 D_{ch}}. \tag{1.8}$$

Thus, we can write the diffusion resistance, Eq. (1.6), as

$$R_{dif}^{(cyl)} = R_{acc} + R_{ws}^{(cyl)}. \tag{1.9}$$

The outline of this paper is as follows. In the following Section 2 we begin with the blocker of a smoothly varying radius, an example of which is shown in Fig. 1a. After that in Section 3 we consider the case of the blocker of a sharply varying radius, namely, a blocking cylinder shown in Fig. 1b. Simulation results supporting our analytical theory are discussed in Section 4. Some concluding remarks are made in the final Section 5.

## 2. Blocker of smoothly varying radius

In this section we derive an expression for the diffusion resistance of a cylindrical channel containing a blocker of a smoothly varying radius, an example of which is shown in Fig. 1a. This is done using the general relation between the diffusion resistance and the translocation probability $P_{tr}$, Eq. (1.4). In finding $P_{tr}$ we take advantage of the fact that when the blocker radius $r(x)$ is a slowly varying function of the coordinate $x$ measured along the channel axis, $|dr(x)/dx| < 1$, one can use an approximate one-dimensional description of the particle dynamics in the channel,[12] in which the particle propagator satisfies the generalized Fick-Jacobs equation.[13]

Let the left and right channel openings be located at $x = 0$ and $x = L$, respectively. The one-dimensional intra-channel propagator (the Green's function), denoted by $G(x, t | x_0)$, $0 < x, x_0 < L$, is the probability density of finding the particle at point $x$ at time $t$, conditional on that this particle (i) was at point $x_0$ at $t = 0$ and (ii) did not escape from the channel during time $t$. The generalized Fick-Jacobs equation for the propagator is

$$\frac{\partial G}{\partial t} = \frac{\partial}{\partial x}\left[A_{ch}(x) D_{ch}(x) \frac{\partial}{\partial x}\left(\frac{G}{A_{ch}(x)}\right)\right], \quad 0 < x < L, \qquad (2.1)$$

where $A_{ch}(x)$ and $D_{ch}(x)$ are the position-dependent channel cross-section area available for diffusing point particles and the intra-channel particle diffusivity. One can find explicit expressions for $D_{ch}(x)$ in papers cited in Ref. 10. At the channel ends the propagator satisfies the radiation boundary conditions that describe the particle escape from the channel[14]

$$A_{ch}(0)D_{ch}(0)\frac{\partial}{\partial x}\left(\frac{G(x,t|x_0)}{A_{ch}(x)}\right)\bigg|_{x=0} = \kappa_0 G(0,t|x_0), \qquad (2.2)$$

$$-A_{ch}(L)D_{ch}(L)\frac{\partial}{\partial x}\left(\frac{G(x,t|x_0)}{A_{ch}(x)}\right)\bigg|_{x=L} = \kappa_0 G(L,t|x_0), \qquad (2.3)$$

where $\kappa_0$ is the particle trapping rate by the channel boundary given by

$$\kappa_0 = \frac{4D_b}{\pi R}. \qquad (2.4)$$

To find $P_{tr}$ consider a steady state with a constant particle flux $J$ injected into the channel near its left opening as shown in Fig. 2. Injected particles escape from the channel through its boundaries. Denoting the fluxes escaping through the left and right boundaries by $J_-$ and $J_+$, respectively, we can write

$$J = J_- + J_+. \qquad (2.5)$$

We use these fluxes to find the translocation probability, which can be written in terms of the fluxes as

$$P_{tr} = J_+/J. \qquad (2.6)$$

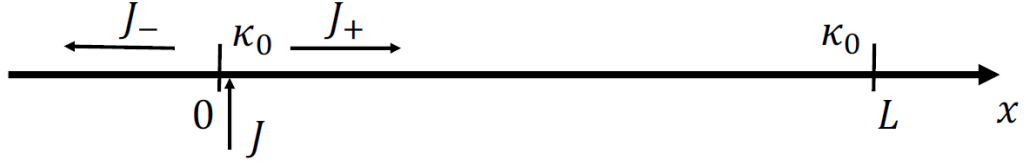

**Fig. 2.** A constant flux $J$ injected into the channel near its left opening and two its components $J_-$ and $J_+$ escaping the channel through its left and right openings, respectively.

Denoting the one-dimensional steady-state concentration of the particles in the channel by $c_1(x)$, we can write the fluxes $J_-$ and $J_+$ in terms of the particle concentrations at the channel ends as

$$J_- = \kappa_0 c_1(0), \qquad J_+ = \kappa_0 c_1(L). \tag{2.7}$$

Inside the channel the concentration $c_1(x)$ satisfies

$$J_+ = -A_{ch}(x) D_{ch}(x) \frac{d}{dx}\left(\frac{c_1(x)}{A_{ch}(x)}\right). \tag{2.8}$$

Dividing both sides of this equation by the product $A_{ch}(x) D_{ch}(x)$ and integrating the resulting equation over $x$ from $0$ to $L$, we obtain

$$J_+ \int_0^L \frac{dx}{A_{ch}(x) D_{ch}(x)} = \frac{c_1(0)}{A_{ch}(0)} - \frac{c_1(L)}{A_{ch}(L)}. \tag{2.9}$$

The areas $A_{ch}(0)$, $A_{ch}(L)$, and $A_{ch}(x)$ entering the above equation are $A_{ch}(0) = A_{ch}(L) = \pi R^2$ and $A_{ch}(x) = \pi(R^2 - r^2(x))$. Using this and Eq. (2.9), we can write the concentration $c_1(0)$ as

$$c_1(0) = c_1(L) + J_+ \int_0^L \frac{dx}{D_{ch}(x)\left(1 - r^2(x)/R^2\right)}$$
$$= J_+ \left[ \frac{1}{\kappa_0} + \int_0^L \frac{dx}{D_{ch}(x)\left(1 - r^2(x)/R^2\right)} \right],$$
(2.10)

where we have used the boundary condition at $x = L$, Eq. (2.7).

The above expression for $c_1(0)$ allows us to find the relation between the fluxes $J_-$ and $J_+$ using the boundary condition at $x=0$, Eq. (2.7). According to this boundary condition we have

$$J_- = \kappa_0 c_1(0) = \alpha J_+,$$
(2.11)

where

$$\alpha = 1 + \kappa_0 \int_0^L \frac{dx}{D_{ch}(x)\left(1 - r^2(x)/R^2\right)}.$$
(2.12)

Substituting $J_-$ in Eq. (2.11) into Eq. (2.5), we obtain

$$J = (1 + \alpha) J_+.$$
(2.13)

This leads to the following expression for the inverse of the translocation probability

$$\frac{1}{P_{tr}} = \frac{J}{J_+} = 1 + \alpha = 2 + \kappa_0 \int_0^L \frac{dx}{D_{ch}(x)\left(1 - r^2(x)/R^2\right)}.$$
(2.14)

Using this in Eq. (1.4), we can write the result in the form similar to that in Eq. (1.9),

$$R_{dif} = R_{acc} + R_{ws},$$
(2.15)

where the access resistance $R_{acc}$ is given by Eq. (1.7), and the intrinsic diffusion resistance $R_{ws}$ of the channel containing the blocker is

$$R_{ws} = \frac{\kappa_0}{4D_b R} \int_0^L \frac{dx}{D_{ch}(x)\left(1 - r^2(x)/R^2\right)}. \tag{2.16}$$

Substituting here the expression for $\kappa_0$, Eq. (2.4), we arrive at

$$R_{ws} = \frac{1}{\pi R^2} \int_0^L \frac{dx}{D_{ch}(x)\left(1 - r^2(x)/R^2\right)} = \int_0^L \frac{dx}{A_{ch}(x) D_{ch}(x)}. \tag{2.17}$$

Expressions in Eqs. (2.15) and (2.17) are the main results of this section. These expressions together with the expression for the access resistance in Eq. (1.7) completely determine the diffusion resistance of a cylindrical channel with a blocker of smoothly varying radius.

As an illustrative example, consider the case where the blocker is formed by two identical cones (see Fig. 1 (a)) of the base radius $R_b$, $R_b < R$, and height $h$, connected by their bases. The blocker is located between points $x = x_- > 0$ and $x = x_+ < L$, $x_+ - x_- = 2h$. Its radius $r(x)$ is given by

$$r(x) = \begin{cases} \lambda(x - x_-), & x_- < x < (x_- + x_+)/2 \\ \lambda(x_+ - x), & (x_- + x_+)/2 < x < x_+ \end{cases}, \tag{2.18}$$

where $\lambda = R_b/h$. The requirement that the blocker radius is a slowly varying function of $x$ imposes a constraint on the parameter $\lambda$, $\lambda < 1$. As follows from Eq. (2.18), the channel cross-section area $A_{ch}(x)$ is

$$A_{ch}(x) = \begin{cases} \pi R^2, & 0 < x < x_-, \ x_+ < x < L \\ \pi\left(R^2 - r^2(x)\right), & x_- < x < x_+ \end{cases}. \tag{2.19}$$

The particle intra-channel diffusivity $D_{ch}(x)$ is a function of $|dr(x)/dx|$.[13b-f, 13h] In the case under consideration $|dr(x)/dx| = const$, and $D_{ch}(x)$ is given by

$$D_{ch}(x) = \begin{cases} D_0, & 0 < x < x_-, \quad x_+ < x < L \\ D_\lambda, & x_- < x < x_+ \end{cases}, \qquad (2.20)$$

where $D_0$ is the particle diffusivity in a cylindrical channel of radius $R$, and $D_\lambda$ is its constant diffusivity in the part of the channel containing the blocker. The simplest expression for $D_\lambda$ was derived by Zwanzig [13b] by the perturbation theory, $D_\lambda = D_0/(1 + \lambda^2/2)$. A more accurate expression was proposed by Reguera and Rubi [13f], $D_\lambda = D_0/\sqrt{1+\lambda^2}$. One can find a detailed discussion of the reduction of the three-dimensional diffusion in a tube of varying radius to the effective one-dimensional description in terms of the entropy potential with the position-dependent diffusivity in Ref. 13h, including various analytical expressions for $D_\lambda$ in Refs. 13b-f, 13h.

Finally, we find the intrinsic diffusion resistance of the cylindrical channel containing the blocker formed by two cones connected by their bases by substituting the above relations into Eq. (2.17) and performing the integration. The result is

$$R_{ws} = \frac{L - 2h}{\pi R^2 D_0} + \frac{h}{\pi R R_b D_\lambda} \ln\left(\frac{R + R_b}{R - R_b}\right). \qquad (2.21)$$

When the cone base radius approaches the channel radius, $R_b \to R$, the blocker completely blocks the channel, and the intrinsic diffusion resistance $R_{ws}$ diverges as $\ln[1/(R-R_b)]$. In the opposite limiting case where the base radius tends to zero this diffusion resistance reduces to $R_{ws}^{cyl}$ in Eq. (1.8).

## 3. Blocker of abruptly changing radius (blocking cylinder)

In Section 2 we derived an expression for the diffusion resistance $R_{dif}$ of a membrane channel containing a blocker of a smooth shape, whose radius $r(x)$ is a slowly varying function of $x$, $|dr(x)/dx| < 1$. The key step in our derivation is the use of an approximate one-dimensional description of the particle diffusive dynamics in the channel. This allowed us to find the translocation probability $P_{tr}$, Eq. (2.14), which after being substituted into Eq. (1.4) led to the expression for $R_{dif}$ in Eq. (2.15) with $R_{acc}$ and $R_{ws}$ given by Eqs. (1.7) and (2.17).

When the blocker radius $r(x)$ changes abruptly, as in the case of a blocking cylinder shown in Fig. 1b, this approach is inapplicable. Here to find $P_{tr}$ we use a different approximate one-dimensional description of the particle dynamics in the channel, which is illustrated in Fig. 3 for the blocking cylinder of radius $a$ and length $l$. We describe particle dynamics as free one-dimensional diffusion both in the channel region containing the blocker, $x_- < x < x_+$, $x_+ = x_- + l$, and outside this region. Particle transitions between the regions are described as trapping by partially absorbing region boundaries. The trapping rates $\kappa$ and $\kappa'$ entering the boundary conditions on the opposite sides of the boundary separating the regions (see Fig. 3) are different.

However, these rates must satisfy the requirement of no net flux across the boundary at equilibrium (detailed balance).

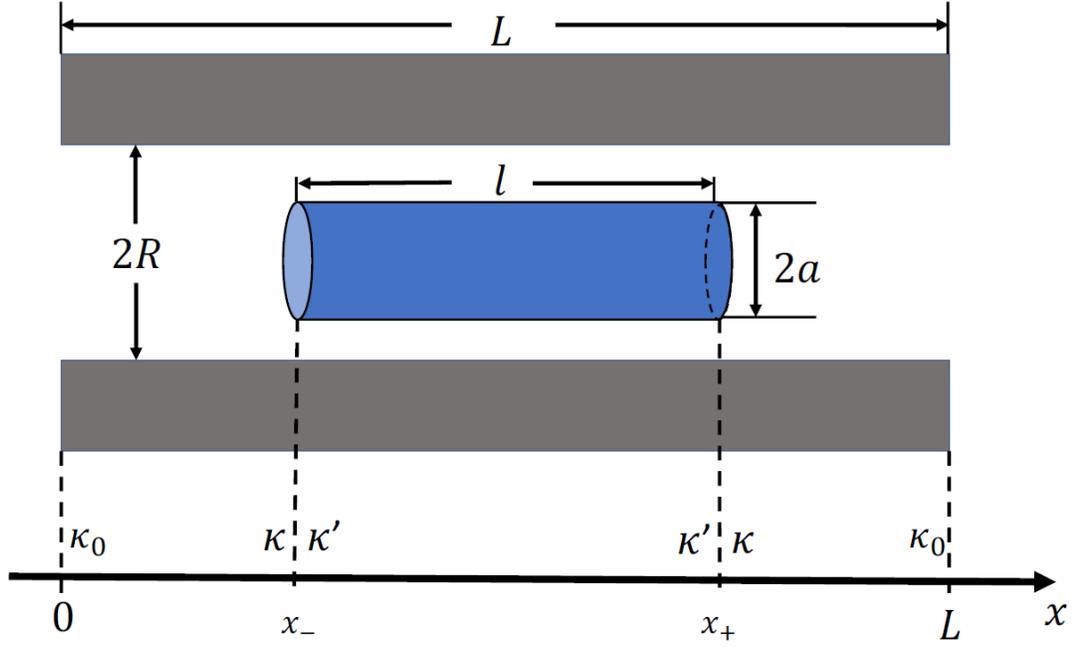

**Fig. 3.** Schematic representation of a cylindrical channel containing a blocking cylinder.

In this description, the one-dimensional particle propagator inside the regions satisfies

$$\frac{\partial G}{\partial t} = \begin{cases} D_{ch} \dfrac{\partial^2 G}{\partial x^2}, & 0 < x < x_-, \quad x_+ < x < L \\ D'_{ch} \dfrac{\partial^2 G}{\partial x^2}, & x_- < x < x_+ \end{cases}, \quad (3.1)$$

where we have assumed that the particle diffusivity in the two regions, $D_{ch}$ and $D'_{ch}$, may be different. To guarantee probability conservation, the propagator must obey the following matching conditions at the region boundaries at points $x = x_-$ and $x = x_+$ (see Fig. 3)

$$D_{ch} \frac{\partial G}{\partial x}\bigg|_{x=x_- -0} = D'_{ch} \frac{\partial G}{\partial x}\bigg|_{x=x_- +0} = \kappa' G\big|_{x=x_- +0} - \kappa G\big|_{x=x_- -0}, \qquad (3.2)$$

$$D'_{ch} \frac{\partial G}{\partial x}\bigg|_{x=x_+ -0} = D_{ch} \frac{\partial G}{\partial x}\bigg|_{x=x_+ +0} = \kappa G\big|_{x=x_+ +0} - \kappa' G\big|_{x=x_+ -0}. \qquad (3.3)$$

Here, the trapping rate $\kappa$ is (see Appendix A)

$$\kappa = \frac{D_{ch}}{R} f(a/R), \qquad f(\zeta) = \frac{3\pi g(\zeta)}{4\zeta^3 \{1 + \ln[1/(1-\zeta)]\}}, \qquad (3.4)$$

where $f(\zeta)$ is the dimensionless function of the dimensionless radius of the blocker $\zeta = a/R$, $0 \leq \zeta \leq 1$, and $g(\zeta)$ is a non-monotonic function given by

$$g(\zeta) = 1 + 0.6\zeta + 2\zeta^2 - 1.5\zeta^3 - 0.8\zeta^{100}. \qquad (3.5)$$

This function is obtained by interpolating our simulation results. The function first slowly increases from 1 at $\zeta = 0$ to about 2 near $\zeta = 0.9$, and then sharply decreases to 4/3 at $\zeta = 1$ (see Fig. 7 in Appendix A). To describe this sharp decrease, we use the term proportional to $\zeta^{100}$ in Eq. (3.5). As the blocker radius approaches the channel radius, $a \to R$ ($\zeta \to 1$), function $f(\zeta)$ tends to zero, and the trapping rate $\kappa$ vanishes, since the particles cannot enter the channel region containing the blocker. In the opposite limit when the blocker radius vanishes, $a$, $\zeta \to 0$, function $f(\zeta)$ diverges, and the trapping rate tends to infinity, as it must be in this limiting case.

We find the trapping rate $\kappa'$ from the requirement of no net flux across the region boundaries at equilibrium. Let $c_{eq}$ be an equilibrium particle concentration in the system. The equilibrium left-to-right unidirectional flux across the region boundary located at $x = x_-$ in our formalism is given by $\kappa \pi R^2 c_{eq}$ (see Eq. (3.2)). The equilibrium unidirectional flux across this boundary in the opposite direction is $\kappa' \pi (R^2 - a^2) c_{eq}$. These two fluxes compensate each other when the trapping rate $\kappa'$ is

$$\kappa' = \frac{\kappa}{1 - a^2/R^2}. \tag{3.6}$$

Finally, as in Section 2, we describe particle escape from the channel to the reservoirs by treating the channel ends as partially absorbing boundaries. The trapping rate $\kappa_0$ entering the boundary conditions at the channel end points $x = 0$ and $x = L$ (see Fig. 3) is given by Eq. (2.4).

The above equations provide a complete one-dimensional description of the particle diffusive dynamics in a cylindrical channel containing a blocking cylinder. In what follows we take advantage of these equations to find the translocation probability $P_{tr}$ which is then used to find the diffusion resistance of the channel. It is worth noting that the one-dimensional description discussed above is applicable on condition that the blocker is located sufficiently far from the channel ends (see Appendix A). Specifically, the distances to both ends cannot be shorter than the channel radius, $x_-, L - x_+ \geq R$.

To find the translocation probability we again consider the steady state with a constant flux $J$ injected into the channel near its left opening at $x=0$ (see Fig. 2). As in Section 2, this flux is the sum of fluxes $J_-$ and $J_+$, Eq. (2.5), which are defined in Eq. (2.7), and the translocation probability $P_{tr}$ is given by the flux ratio, Eq. (2.6). The one-dimensional particle concentration $c_1(x)$ in different regions of the channel with the blocking cylinder satisfies

$$J_+ = \begin{cases} -D_{ch}\dfrac{dc_1(x)}{dx}, & 0 < x < x_-, \quad x_+ < x < L \\ -D'_{ch}\dfrac{dc_1(x)}{dx}, & x_- < x < x_+ \end{cases} \quad (3.7)$$

The matching conditions at the region boundaries are

$$J_+ = \kappa c_1(x_- - 0) - \kappa' c_1(x_- + 0) = \kappa' c_1(x_+ - 0) - \kappa c_1(x_+ + 0). \quad (3.8)$$

Solving the above equations subject to the boundary condition $c_1(L) = J_+/\kappa_0$ (see Eq. (2.7)), we obtain $c_1(0)$ as a function of $J_+$

$$c_1(0) = J_+\left(\dfrac{2}{\kappa} + \dfrac{1}{\kappa_0} + \dfrac{L-l}{D_{ch}} + \dfrac{\kappa'}{\kappa}\dfrac{l}{D'_{ch}}\right). \quad (3.9)$$

Using the boundary condition in Eq. (2.7) at $x=0$ and Eq. (2.5), we arrive at the relation between the fluxes $J$ and $J_+$ in Eq. (2.13) with $\alpha$ given by

$$\alpha = 1 + \kappa_0\left(\dfrac{2}{\kappa} + \dfrac{L-l}{D_{ch}} + \dfrac{\kappa'}{\kappa}\dfrac{l}{D'_{ch}}\right). \quad (3.10)$$

This leads to the following expression for the inverse of the translocation probability (cf. Eq. (2.14))

$$\frac{1}{P_{tr}} = \frac{J}{J_+} = 1+\alpha = 2+\kappa_0\left(\frac{2}{\kappa}+\frac{L-l}{D_{ch}}+\frac{\kappa'}{\kappa}\frac{l}{D'_{ch}}\right). \qquad (3.11)$$

Substituting this into Eq. (1.4), we arrive at the expression for the diffusion resistance of the channel with the blocker which can be written as

$$R_{dif} = \frac{1}{4RD_b P_{tr}} = R_{acc} + R_{ws}, \qquad (3.12)$$

where $R_{acc}$ is the access resistance given in Eq. (1.7), and $R_{ws}$ is the intrinsic diffusion resistance of the channel with the blocking cylinder given by

$$R_{ws} = \frac{\kappa_0}{4RD_b}\left(\frac{2}{\kappa}+\frac{L-l}{D_{ch}}+\frac{\kappa'}{\kappa}\frac{l}{D'_{ch}}\right). \qquad (3.13)$$

As follows from Eqs. (2.4) and (3.6), the two ratios entering the above equation are

$$\frac{\kappa_0}{4RD_b} = \frac{1}{\pi R^2}, \qquad \frac{\kappa'}{\kappa} = \frac{R^2}{R^2 - a^2}. \qquad (3.14)$$

These allow us to write $R_{ws}$ in Eq. (3.13) as

$$R_{ws} = \frac{2}{\pi R^2 \kappa} + \frac{L-l}{\pi R^2 D_{ch}} + \frac{l}{\pi(R^2-a^2)D'_{ch}}$$
$$= \frac{2M(a/R)}{\pi R D_{ch}} + \frac{L-l}{\pi R^2 D_{ch}} + \frac{l}{\pi(R^2-a^2)D'_{ch}}, \qquad (3.15)$$

where we have used the expression for $\kappa$ in Eq. (3.4) and introduced function $M(\zeta)$, $\zeta = a/R$, defined by

$$M(\zeta) = \frac{1}{f(\zeta)} = \frac{4\zeta^3\{1+\ln[1/(1-\zeta)]\}}{3\pi g(\zeta)}. \qquad (3.16)$$

As $\zeta$ increases from zero to unity, this function monotonically increases from zero to infinity, as illustrated in Fig. 4.

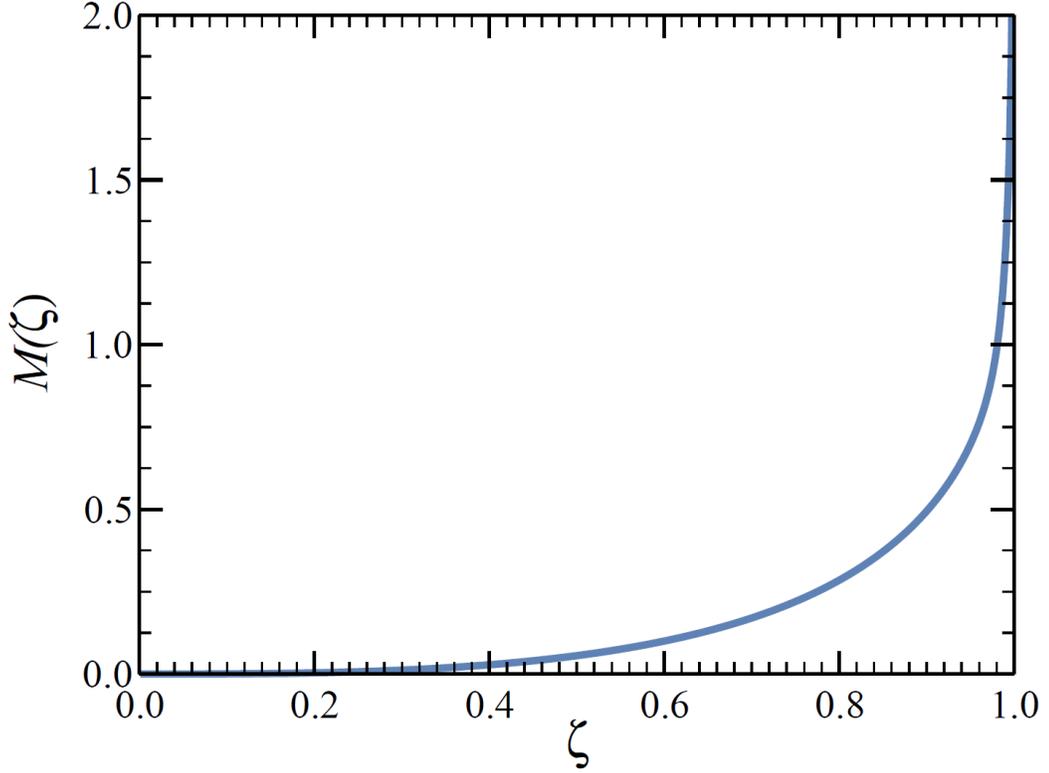

**Fig. 4.** Function $M(\zeta)$, Eq. (3.16), $\zeta = a/R$.

When $l \to 0$, the blocking cylinder becomes an infinitely thin disk-blocker. In this case $R_{ws}$ in Eq. (3.15) is the sum of the intrinsic diffusion resistance of the cylindrical channel without a blocker, $R_{ws}^{(cyl)}$ in Eq. (1.8), and the additional diffusion resistance of the infinitely thin disk-blocker (*disk*) of radius $a$. Denoting this additional diffusion resistance by $R_{disk}$,

$$R_{disk} = \frac{2}{\pi R D_{ch}} M(a/R), \quad (3.17)$$

we can write Eq. (3.15) as

$$R_{ws} = \frac{L-l}{\pi R^2 D_{ch}} + \frac{l}{\pi (R^2 - a^2) D'_{ch}} + R_{disk}. \quad (3.18)$$

As the blocker radius vanishes, $a, \zeta \to 0$, $D'_{ch}$ becomes equal to $D_{ch}$ and $R_{ws}$ reduces to $R_{ws}^{(cyl)}$ in Eq. (1.8). In the opposite limit, where the blocker radius approaches to that of the channel, $a \to R$, $\zeta \to 1$, both $R_{ws}$ and $R_{dif}$ diverge as $1/(R-a)$ since diffusing point particles cannot path through the channel completely blocked by the blocker.

Note that the first two terms in Eq. (3.18) describe contributions to $R_{ws}$ of the channel region containing the blocker and the rest of the channel. (They can be obtained by using Eq. (2.17).) The last term $R_{disk}$ is the contribution due to the particle transitions between different regions of the channel. Naturally, this term can be neglected for long blockers. However, it is important for short blockers whose lengths satisfy $l < 2RM(\zeta)(1-\zeta^2) D'_{ch}/D_{ch}$. This inequality shows that the range of $l$, where the contribution of $R_{disk}$ to $R_{ws}$ is important, is sensitive to the blocker radius; the range vanishes as $\zeta \to 0$ or 1 and reaches its maximum at intermediate values of $\zeta$, i.e., at intermediate values of $a$, which are not too small and not too close to the channel radius $R$.

To summarize, main results of this section are the expressions in Eqs. (3.12), (3.17), and (3.18), which together with the expression for the access resistance, Eq.

(1.7), completely determine the diffusion resistance of a cylindrical channel containing a blocking cylinder located not too close to the channel ends.

## 4. Simulation results

This section discusses a simulation test of our analytical expression for the intrinsic diffusion resistance $R_{ws}$ in Eq. (3.18). Of course, it would be great to verify our analytical results by three-dimensional Brownian dynamics simulations. To determine $R_{dif}$ one has to simulate particle transport between the two reservoirs. This involves the diffusion of particles in the reservoirs, their transitions between the pore and the reservoirs, and their diffusion in the pore. Unfortunately, this would require an unrealistic amount of computational resources. To circumvent this difficulty, we propose an alternative approach, namely, we first derive the relation between $R_{ws}$ and the mean first passage times (MFPTs) of the diffusing particles between the channel ends. Then we use this relation to verify our approximate theory. Let $\tau_{FP}(0 \to L)$ be the MFPT of a diffusing particle from the reflecting boundary of the interval, located at $x=0$, to its absorbing end point, located at $x=L$. Its counterpart for the transition in the opposite direction is denoted by $\tau_{FP}(L \to 0)$. These two MFPTs allows us to find $R_{ws}$ of the channel containing the blocker using the following relationship

$$R_{ws} = \frac{\tau_{FP}(0 \to L) + \tau_{FP}(L \to 0)}{V_{ch} - V_{bl}}, \qquad (4.1)$$

where $V_{ch}$ and $V_{bl}$ are the channel and blocker volumes, respectively. This relationship is derived below in the framework of the approximate one-dimensional description of the particle diffusive dynamics in the channel discussed in Section 3. It is worth noting that an analogous relationship between $R_{ws}$ and the sum of the MFPTs between the channel ends can be derived for a blocker of a smoothly varying radius discussed in Section 2.

To find the MFPT $\tau_{FP}(0 \to L)$ consider a steady state where a constant flux $J$ is injected into the interval near its reflecting left boundary located at $x=0$. The particles escape from the interval through its absorbing right boundary located at $x=L$ (see Fig. 5a). The steady-state particle concentration $c_1(x)$, as before, satisfies Eq. (3.7) and the matching conditions in Eq. (3.8) at points $x=x_-$ and $x=x_+$, with $J_+$ replaced by $J$. However, the boundary conditions at the ends of the interval are now different, since the left and right ends of the interval are reflecting and absorbing boundaries, respectively. We determine $c_1(x)$ by solving these equations and then find $\tau_{FP}(0 \to L)$ by means of the relationship

$$\tau_{FP}(0 \to L) = \frac{1}{J}\int_0^L c_1(x)dx. \tag{4.2}$$

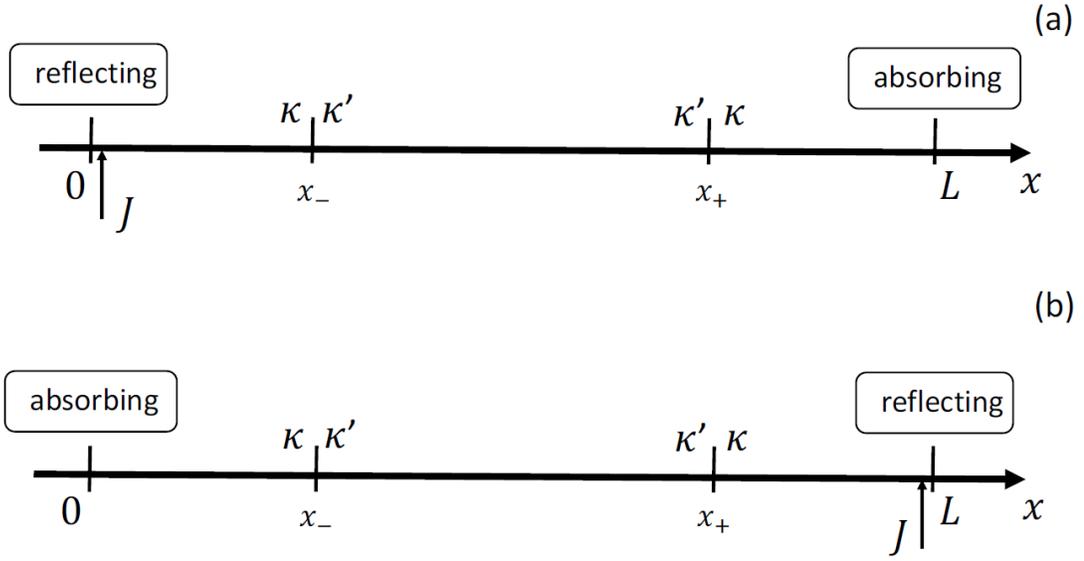

**Fig. 5 (a), (b).** Constant fluxes injected near the reflecting end of the interval and flowing to its absorbing end: from left to right in panel (a) and in the opposite direction in panel (b).

One can check that $c_1(x)$ is given by

$$c_1(x) = \begin{cases} c_1(x_- - 0) + J \dfrac{x_- - x}{D_{ch}}, & 0 < x < x_- \\ c_1(x_+ - 0) + J \dfrac{x_+ - x}{D'_{ch}}, & x_- < x < x_+ \\ J \dfrac{L - x}{D_{ch}}, & x_+ < x < L \end{cases} \quad (4.3)$$

where the concentrations $c_1(x_- - 0)$ and $c_1(x_+ - 0)$ are

$$c_1(x_- - 0) = J\left(\frac{2}{\kappa} + \frac{L - x_+}{D_{ch}} + \frac{\kappa'}{\kappa}\frac{l}{D'_{ch}}\right) \quad (4.4)$$

and

$$c_1(x_+ - 0) = J\left(\frac{1}{\kappa'} + \frac{\kappa}{\kappa'}\frac{L - x_+}{D_{ch}}\right). \quad (4.5)$$

Substituting the above expression for $c_1(x)$ into Eq. (4.2) and performing the integration, we arrive at

$$\tau_{FP}(0 \to L) = \frac{1}{2D_{ch}}\left[x_-^2 + (L-x_+)^2\right] + \frac{1}{2D'_{ch}}l^2 +$$
$$\frac{L-x_+}{D_{ch}}\left[x_- + l(1-a^2/R^2)\right] + \frac{lx_-}{D'_{ch}(1-a^2/R^2)} + \frac{1}{\kappa}\left[2x_- + l(1-a^2/R^2)\right], \quad (4.6)$$

where we have used the relation between $\kappa'$ and $\kappa$ in Eq. (3.6).

To find the MFPT $\tau_{FP}(L \to 0)$ one has to repeat similar calculation for the steady state shown in Fig. 5b, where a constant flux $J$ is injected into the interval near its reflecting right boundary located at $x = L$ and the particles are trapped at $x = 0$ by the absorbing left end of the interval. This leads to the following expression for the MFPT $\tau_{FP}(L \to 0)$,

$$\tau_{FP}(L \to 0) = \frac{1}{2D_{ch}}\left[x_-^2 + (L-x_+)^2\right] + \frac{1}{2D'_{ch}}l^2 +$$
$$\frac{x_-}{D_{ch}}\left[L-x_+ + l(1-a^2/R^2)\right] + \frac{l(L-x_+)}{D'_{ch}(1-a^2/R^2)} + \frac{1}{\kappa}\left[2(L-x_+) + l(1-a^2/R^2)\right]. \quad (4.7)$$

Using Eq. (4.6) and (4.7), one can check that the sum of the MFPTs is equal to $R_{ws}$ multiplied by the volume difference,

$$\tau_{FP}(0 \to L) + \tau_{FP}(L \to 0) = R_{ws}(V_{ch} - V_{bl}), \quad (4.8)$$

where the channel and blocker volumes, respectively, are

$$V_{ch} = \pi R^2 L, \qquad V_{bl} = \pi a^2 l. \quad (4.9)$$

Dividing both sides of Eq. (4.8) by the volume difference $V_{ch} - V_{bl}$, we recover the relationship in Eq. (4.1), which is used to test the expression for $R_{ws}$ in Eq. (3.18) predicted by the theory.

To this end we obtained the MFPTs between the channel ends from three-dimensional Brownian dynamics simulations as described in Appendix B. The blocking cylinder was always placed in the center of the channel. Under such conditions the MFPTs between the channel ends are equal, $\tau_{FP}(0 \to L) = \tau_{FP}(L \to 0)$, and their sum is $2\tau_{FP}(0 \to L)$.

To check the theory, we consider the ratio of the sum of the MFPTs obtained from our simulations to the product of $R_{ws}$ in Eq. (3.18) and the volume difference $V_{ch} - V_{bl}$. The theory predicts that this ratio must be 1 (see Eq. (4.1)). In our simulations we tested this prediction for the channel of length $L = 4R$. The blocker length $l$ and radius $a$, respectively, were $l/R = 0$ (disk), 1, and 2, and $a/R = 1/4$, 1/2, and 3/4. In addition, we ignore variation of the intra-channel diffusivity and take $D'_{ch} = D_{ch}$. The ratios of the sum of the MFPTs to the product of $R_{ws}$ in Eq. (3.18) and the volume difference $V_{ch} - V_{bl}$, for our set of the blocker parameters are summarized in Table 1. One can see a good agreement between the theoretical predictions and the simulation results: the relative error does not exceed 2%.

|       | l/R    |        |        |
|-------|--------|--------|--------|
| *a/R* | **0**  | **1**  | **2**  |
| **0.25** | 0.9999 | 1.0115 | 1.0022 |
| **0.50** | 0.9973 | 0.9971 | 1.0091 |
| **0.75** | 1.0098 | 1.0004 | 1.0029 |

**Table 1.** The ratio $\left(\tau_{FP}(0 \to L) + \tau_{FP}(L \to 0)\right)/\left[R_{ws}(V_{ch} - V_{bl})\right]$ as a function of the blocker length and radius normalized to the channel radius. The MFPTs were obtained from the three-dimensional Brownian dynamics simulations. The intrinsic diffusion resistance $R_{ws}$ and the volumes $V_{ch}$ and $V_{bl}$ are given in Eqs. (3.18) and (4.9), respectively.

## 5. Concluding remarks

This work deals with the steady-state flux of small solutes driven by their concentration difference in two reservoirs connected by a cylindrical nanopore. The focus is on how the presence of a blocker – large analyte molecule or molecular aggregate creating a static obstacle in the nanopore – affects the flux. It is assumed that the blocker is cylindrically symmetric, and its axis coincides with the axis of the nanopore. We quantify the effect in terms of the channel diffusion resistance which is the ratio of the solute concentration difference in the reservoirs to the flux (see Eq. (1.1)). Our main results are the expressions for the diffusion resistance of the nanopore containing a blocker of a slowly varying radius, Eqs. (2.15) and (2.17), or a blocking cylinder, Eqs. (3.12), (3.17), and (3.18). Combined with Eq. (1.1), these expressions show how the presence of the blocker affects the flux. Specifically, they predict how the effect depends on the blocker shape and size.

In both cases the key step in our approach is an approximate reduction of the initial three-dimensional diffusion problem to an equivalent one-dimensional one which can be solved with relative ease. However, the dimensionality reduction in the two cases is performed differently, leading to different equivalent one-dimensional descriptions of the solute dynamics in the nanopore. In the case of a blocker of slowly varying radius this description is formulated in terms of the generalized Fick-Jacobs equation proposed for diffusion in tubes and channels of smoothly varying cross section. In contrast, in the case of the blocking cylinder, the effective one-dimensional dynamics is free diffusion both in the nanopore region containing the blocker and outside this region, Eq. (3.1). The major challenge here is how to treat the transitions between these regions. We describe them by the matching conditions in Eqs. (3.2) and (3.3), which contain the trapping rates $\kappa$ and $\kappa'$. The former, given in Eq. (3.4), is obtained using the boundary homogenization approach to the trapping problem, as discussed in Appendix A. The latter is determined from the former using the detailed balance condition, Eq. (3.6). To test the accuracy of our approach in the case of the blocking cylinder, we compare our theoretical predictions with the corresponding data obtained from three-dimensional Brownian dynamics simulations. The results of this comparison, summarized in Table 1, demonstrate good agreement between the two.

It is interesting that there is a universal relation between the increase of the channel diffusion resistance due to the presence of the blocker and the blocker volume $V_{bl}$, which works in both cases when the blocker is long and narrow. Using Eqs. (2.17) and (3.18), one can check that the increase in the diffusion resistance is given by the ratio $V_{bl}/(\pi^2 R^4 D_{ch})$. This is equivalent to the replacement of the initial cylindrical channel of length $L$ with the blocker by an effective longer empty channel of the same radius and width $L + V_{bl}/(\pi R^2)$.

Finally, we note that the expressions derived for the diffusion resistance establish a connection between the hydrodynamic description of transport through the nanopore in terms of $R_{dif}$ and mesoscopic description of the solute dynamics in the channel/nanopore.

**Acknowledgements**

This study was partially supported by the Intramural Research Program of the NIH, *Eunice Kennedy Shriver* National Institute of Child Health and Human Development. LD was partially supported by CONACyT under grant Frontiers Science No. 51476. ATS thanks Paul A. Martin for many illuminating discussions.

**Data availability statement.** The data that support the findings of this study are available from the corresponding author upon reasonable request.

**Appendix A. Boundary trapping rate**

In this Appendix we obtain the boundary trapping rate $\kappa$, Eq. (3.4), that describes entrance of diffusing point particles in the channel region containing the blocking cylinder of radius $a$ (see Fig. 3). To this end consider point particles diffusing in a straight horizontal cylinder of radius $R$ and length $b$, $b \gg R$, with reflecting side wall, as shown in Fig. 6. The particle concentration at the right face of the cylinder, located at $x=b$, is kept constant. The left face of the cylinder, located at $x=0$, contains a reflecting circular disk of radius $a$, $a \leq R$, in its center. The rest of this face is the absorbing boundary that instantly traps diffusing particles at the first contact.

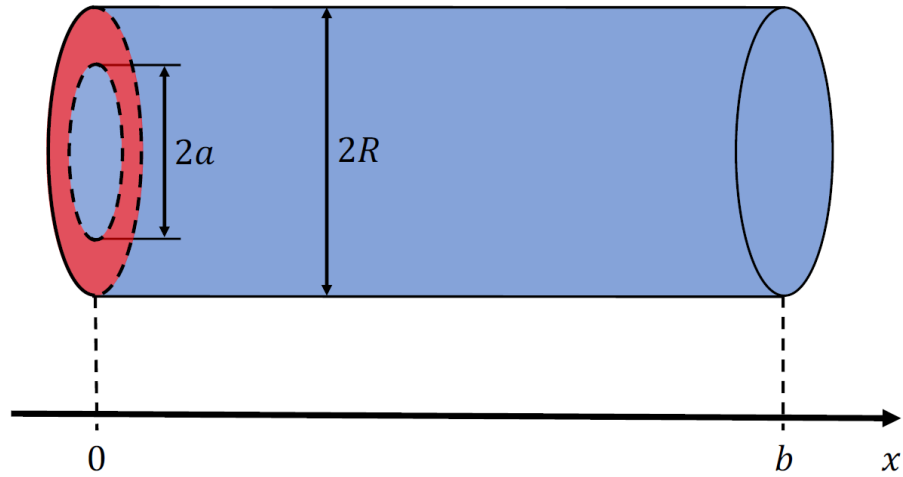

**Fig. 6.** Cylinder of radius $R$ and length $b$ with reflecting side wall. The left face of the cylinder is absorbing except its central part covered by a reflecting circular disk of radius $a$. The absorbing part of this face is colored in red.

The entire system is in steady state with a constant flux $J$ flowing from right to left. Although the flux density $j(x)$ is highly non-uniform near the left face of the cylinder, sufficiently far from this face the flux density is uniform and given by $j = J/(\pi R^2)$. With this in mind, we can replace the inhomogeneous boundary conditions on the left face by a homogeneous one which is partially absorbing with the trapping rate $\kappa(a/R)$ chosen so that to reproduce the same steady-state flux $J$. This is the so-called boundary homogenization (BH).

We assume that when BH is applicable, the effective homogeneous boundary conditions can be used to analyze other diffusion problems in the cylinder, which after the BH become essentially one-dimensional. For example, we can find the mean lifetime of a particle diffusing in the cylinder with the reflecting right face, which is the mean first-passage time of the particle to the absorbing part of the cylinder left face. In the one-dimensional description this is the mean lifetime of a particle diffusing on an interval of length $b$ terminated by reflecting and partially absorbing end points.

The situation where the particle starting point is uniformly distributed over the left face of the cylinder, in the one-dimensional description corresponds to the case of the particle starting from the partially absorbing boundary of the interval. The mean lifetime of this particle, denoted by $\tau_{1D}(a,R,b)$, is given by[15]

$$\tau_{1D}(a,R,b) = \frac{b}{\kappa(a/R)}. \qquad (A.1)$$

We take advantage of this relation to find $\kappa(a/R)$ from the mean particle lifetime obtained from Brownian dynamics simulations. Denoting this mean lifetime by $\tau_{sim}(a,R,b)$, we can write

$$\kappa(a/R) = \lim_{b \to \infty} \frac{b}{\tau_{sim}(a,R,b)}. \tag{A.2}$$

We run the simulations in three dimensions for the following set of parameters of the cylinder: $a/R = 0.1, 0.2, 0.3, 0.4, 0.5, 0.6, 0.7, 0.8, 0.9, 0.95, 0.97, 0.98, 0.99$ and $b/R = 0.25, 0.5, 0.75, 1.0, 1.25, 1.5, 1.75, 2.0$. The simulation results show that the ratio $b/\tau_{sim}(a,R,b)$ reaches its plateau value $\kappa(a/R)$ when the cylinder length equals or exceeds its radius, $b/R \geq 1$. This is true for all values of the disk radius $a$ used in our simulations.

The values of $\kappa(a/R)$ found from our simulations are used to construct an approximate formula for the trapping rate, Eq. (3.4). In doing so, we take advantage of the circumstance that the exact asymptotic behavior of $\kappa(a/R)$ can be determined using some exact asymptotic results derived in hydrodynamics.[16] Specifically, as $a/R$ approaches 0 and 1, $\kappa(a/R)$, respectively, tends to infinity and zero as

$$\kappa(a/R) = \frac{D_{ch}}{R} \times \begin{cases} (3\pi/4)(R/a)^3, & a \to 0 \\ \pi/\ln[1/(1-a/R)], & a \to R \end{cases}. \tag{A.3}$$

By taking advantage of the above asymptotic behavior, we can write the trapping rate $\kappa(\zeta)$, $\zeta = a/R$, in the form given in Eq. (3.4) with function $g(\zeta)$ obtained using our simulation results. This function monotonically increases with $\zeta$ from 1 at $\zeta = 0$ to its maximum value about 2 at $\zeta \simeq 0.9$ and then sharply decreases to its limiting value 4/3 at $\zeta =1$, as shown in Fig. 7. The expression in Eq. (3.5) interpolates function $g(\zeta)$ with the relative error less than 4% in the entire range of its argument, that is $0 \le \zeta \le 1$.

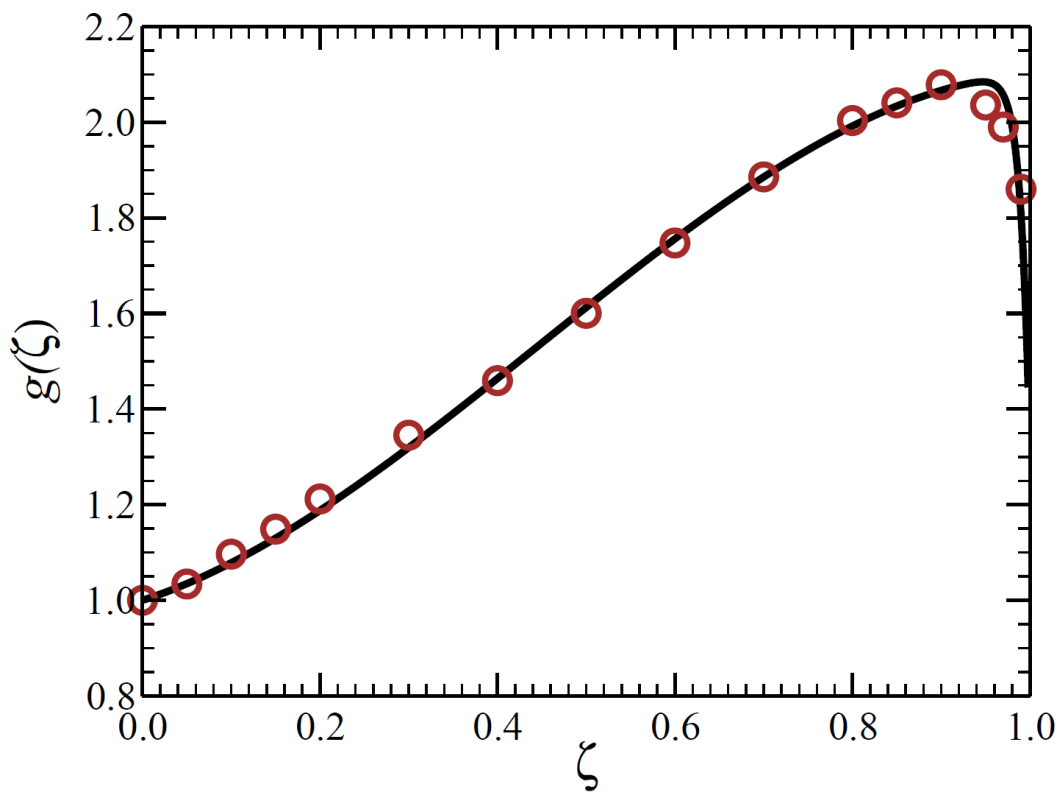

**Fig. 7.** Interpolating function $g(\zeta)$, Eq. (3.5), $\zeta = a/R$, (solid curve) and the values of this function obtained from our Brownian dynamics simulations (circles).

**Appendix B. Simulation details**

In our three-dimensional Brownian dynamics simulations we determine the MFPT $\tau_{FP}(L \to 0)$ by running $N = 2.5 \times 10^4$ trajectories of overdamped Brownian particles. The trajectories start from the reflecting channel boundary located at $x = L$ and are terminated as soon as they touch the absorbing boundary located at $x = 0$ for the first time. The initial distribution of the starting point is uniform over the reflecting boundary. The MFPT is obtained by averaging of the FPT's of individual trajectories $t_n$,

$$\tau_{FP}(L \to 0) = \frac{1}{N} \sum_{n=1}^{N} t_n. \tag{B.1}$$

In our simulations we use dimensionless variables. To this end we choose the channel radius $R$ as a unit of length and take the intra-channel diffusivity equal to unity, $D_{ch} = 1$. Then the ratio $R^2/D_{ch}$ provides a unit for measuring time. In addition, we ignore variation of the intra-channel diffusivity and take $D'_{ch} = D_{ch}$. In dimensionless variables the overdamped Langevin equation of motion of a Brownian particle is

$$\frac{dx_i}{dt} = \sqrt{2}\xi_i, \qquad i = x, y, z, \tag{B.2}$$

where $\xi_i(t)$ is the zero-mean $\delta$-correlated Gaussian white noise, $\langle \xi_i(t) \rangle = 0$, with the correlation function $\langle \xi_i(t)\xi_j(t') \rangle = \delta_{ij}\delta(t-t')$, $i, j = x, y, z$; here the angular brackets $\langle ... \rangle$ denote the averaging over realizations of the random force. We integrate Eq. (B.2) numerically with the time step $\Delta t = 10^{-7}$, so that the length step $\sqrt{2\Delta t} = 1$. This is done with reflecting boundary conditions on the channel wall and on the surface of the blocker. In simulations we modeled these boundary conditions as classical elastic collisions of the particle with the surfaces.

**Figure captions**

**Fig. 1 (a), (b).** Schematic representation of a cylindrical channel with a blocker formed by two cones connected by their bases (panel (a)) and with a blocking cylinder (panel (b)).

**Fig. 2.** A constant flux $J$ injected into the channel near its left opening and two its components $J_-$ and $J_+$ escaping the channel through its left and right openings, respectively.

**Fig. 3.** Schematic representation of a cylindrical channel containing a blocking cylinder.

**Fig. 4.** Function $M(\zeta)$, Eq. (3.16), $\zeta = a/R$.

**Fig. 5 (a), (b).** Constant fluxes injected near the reflecting end of the interval and flowing to its absorbing end: from left to right in panel (a) and in the opposite direction in panel (b).

**Fig. 6.** Cylinder of radius $R$ and length $b$ with reflecting side wall. The left face of the cylinder is absorbing expect its central part covered by a reflecting circular disk of radius $a$. The absorbing part of this face is colored in red.

**Fig. 7.** Interpolating function $g(\zeta)$, Eq. (3.5), $\zeta = a/R$, (solid curve) and the values of this function obtained from our Brownian dynamics simulations (circles).

# References

1. (a) Butler, T. Z.; Pavlenok, M.; Derrington, I. M.; Niederweis, M.; Gundlach, J. H., Single-molecule DNA detection with an engineered MspA protein nanopore. *P Natl Acad Sci USA* **2008,** *105* (52), 20647-20652; (b) Deamer, D.; Akeson, M.; Branton, D., Three decades of nanopore sequencing. *Nat Biotechnol* **2016,** *34* (5), 518-524; (c) Restrepo-Perez, L.; Joo, C.; Dekker, C., Paving the way to single-molecule protein sequencing. *Nat Nanotechnol* **2018,** *13* (9), 786-796; (d) Callahan, N.; Tullman, J.; Kelman, Z.; Marino, J., Strategies for Development of a Next-Generation Protein Sequencing Platform. *Trends Biochem Sci* **2020,** *45* (1), 76-89; (e) Ouldali, H.; Sarthak, K.; Ensslen, T.; Piguet, F.; Manivet, P.; Pelta, J.; Behrends, J. C.; Aksimentiev, A.; Oukhaled, A., Electrical recognition of the twenty proteinogenic amino acids using an aerolysin nanopore. *Nat Biotechnol* **2020,** *38* (2), 176-181.
2. (a) Thakur, A. K.; Movileanu, L., Real-time measurement of protein-protein interactions at single-molecule resolution using a biological nanopore. *Nat Biotechnol* **2019,** *37* (1), 96-+; (b) Hoogerheide, D. P.; Gurnev, P. A.; Rostovtseva, T. K.; Bezrukov, S. M., Voltage-activated complexation of alpha-synuclein with three diverse beta-barrel channels: VDAC, MspA, and alpha-hemolysin. *Proteomics* **2021**, 2100060.
3. (a) Nivala, J.; Marks, D. B.; Akeson, M., Unfoldase-mediated protein translocation through an alpha-hemolysin nanopore. *Nat Biotechnol* **2013,** *31* (3), 247-250; (b) Willems, K.; Van Meervelt, V.; Wloka, C.; Maglia, G., Single-molecule nanopore enzymology. *Philos T R Soc B* **2017,** *372* (1726); (c) Hoogerheide, D.; Gurnev, P.; Rostovtseva, T.; Bezrukov, S., Effect of a post-translational modification mimic on protein translocation through a nanopore. *Nanoscale* **2020,** *12*, 11070.
4. (a) Reiner, J. E.; Kasianowicz, J. J.; Nablo, B. J.; Robertson, J. W. F., Theory for polymer analysis using nanopore-based single-molecule mass spectrometry. *P Natl Acad Sci USA* **2010,** *107* (27), 12080-12085; (b) Kowalczyk, S. W.; Grosberg, A. Y.; Rabin, Y.; Dekker, C., Modeling the conductance and DNA blockade of solid-state nanopores. *Nanotechnology* **2011,** *22* (31); (c) Gurnev, P. A.; Rostovtseva, T. K.; Bezrukov, S. M., Tubulin-blocked state of VDAC studied by polymer and ATP partitioning. *Febs Lett* **2011,** *585* (14), 2363-2366; (d) Gurnev, P. A.; Stanley, C. B.; Aksoyoglu, M. A.; Hong, K. L.; Parsegian, V. A.; Bezrukov, S. M., Poly(ethylene glycol)s in Semidilute Regime: Radius of Gyration in the Bulk and Partitioning into a Nanopore. *Macromolecules* **2017,** *50* (6), 2477-2483; (e) Berezhkovskii, A. M.; Bezrukov, S. M., Capturing single molecules by nanopores: measured times and thermodynamics. *Phys Chem Chem Phys* **2021,** *23* (2), 1610-1615.
5. (a) Aguilella-Arzo, M.; Andrio, A.; Aguilella, V. M.; Alcaraz, A., Dielectric saturation of water in a membrane protein channel. *Phys Chem Chem Phys* **2009,** *11* (2), 358-365; (b) Tsuruta, T., On the Role of Water Molecules in the Interface between Biological Systems and Polymers. *J Biomat Sci-Polym E* **2010,** *21* (14), 1831-1848; (c) Ostmeyer, J.; Chakrapani, S.; Pan, A. C.; Perozo, E.; Roux, B., Recovery from slow inactivation in K+ channels is controlled by water molecules. *Nature* **2013,** *501* (7465).
6. (a) Howorka, S.; Siwy, Z., Nanopore analytics: sensing of single molecules. *Chem Soc Rev* **2009,** *38* (8), 2360-2384; (b) Kowalczyk, S. W.; Kapinos, L.; Blosser, T. R.; Magalhaes, T.; van Nies, P.; Lim, R. Y. H.; Dekker, C., Single-molecule transport across an individual biomimetic nuclear pore complex. *Nat Nanotechnol* **2011,** *6* (7), 433-438; (c) Hoogerheide, D. P.; Albertorio, F.; Golovchenko, J. A., Escape of DNA from a Weakly Biased Thin Nanopore: Experimental Evidence for a Universal Diffusive Behavior. *Phys Rev Lett* **2013,** *111* (24).
7. Skvortsov, A. T.; Dagdug, L.; Berezhkovskii, A. M.; MacGillivray, I. R.; Bezrukov, S. M., Evaluating diffusion resistance of a constriction in a membrane channel by the method of boundary homogenization. *Phys Rev E* **2021,** *103* (1).

8. Mierle, G., The Effect of Cell-Size and Shape on the Resistance of Unstirred Layers to Solute Diffusion. *Biochim Biophys Acta* **1985,** *812* (3), 835-839.
9. (a) Hill, T. L., Effect of Rotation on Diffusion-Controlled Rate of Ligand-Protein Association. *P Natl Acad Sci USA* **1975,** *72* (12), 4918-4922; (b) Berg, H. C.; Purcell, E. M., Physics of Chemoreception. *Biophys J* **1977,** *20* (2), 193-219.
10. Berezhkovskii, A. M.; Pustovoit, M. A.; Bezrukov, S. M., Channel-facilitated membrane transport: Transit probability and interaction with the channel. *J Chem Phys* **2002,** *116* (22), 9952-9956.
11. Hall, J. E., Access Resistance of a Small Circular Pore. *J Gen Physiol* **1975,** *66* (4), 531-532.
12. Berezhkovskii, A. M.; Pustovoit, M. A.; Bezrukov, S. M., Diffusion in a tube of varying cross section: Numerical study of reduction to effective one-dimensional description. *J Chem Phys* **2007,** *126* (13), 134706
13. (a) Jacobs, M. H., *Diffusion Processes*. Springer: New York, 1967; (b) Zwanzig, R., Diffusion Past an Entropy Barrier. *J Phys Chem-Us* **1992,** *96* (10), 3926-3930; (c) Kalinay, P.; Percus, J. K., Extended fick-jacobs equation: Variational approach. *Phys Rev E* **2005,** *72* (6), 061203; (d) Kalinay, P.; Percus, J. K., Corrections to the Fick-Jacobs equation. *Phys Rev E* **2006,** *74* (4), 041203; (e) Kalinay, P.; Percus, J. K., Approximations of the generalized Fick-Jacobs equation. *Phys Rev E* **2008,** *78* (2), 021103; (f) Reguera, D.; Rubi, J. M., Kinetic equations for diffusion in the presence of entropic barriers. *Phys Rev E* **2001,** *64* (6), 061106; (g) Berezhkovskii, A. M.; Szabo, A., Time scale separation leads to position-dependent diffusion along a slow coordinate. *J Chem Phys* **2011,** *135* (7), 074108; (h) Berezhkovskii, A. M.; Bezrukov, S. M., On the applicability of entropy potentials in transport problems. *Eur Phys J-Spec Top* **2014,** *223* (14), 3063-3077.
14. (a) Bezrukov, S. M.; Berezhkovskii, A. M.; Pustovoit, M. A.; Szabo, A., Particle number fluctuations in a membrane channel. *J Chem Phys* **2000,** *113* (18), 8206-8211; (b) Berezhkovskii, A. M.; Szabo, A.; Zhou, H. X., Diffusion-influenced ligand binding to buried sites in macromolecules and transmembrane channels. *J Chem Phys* **2011,** *135* (7), 075103.
15. Bicout, D. J.; Szabo, A., First passage times, correlation functions, and reaction rates. *J Chem Phys* **1997,** *106* (24), 10292-10298.
16. (a) Sherwood, J. D.; Stone, H. A., Added mass of a disc accelerating within a pipe. *Phys Fluids* **1997,** *9* (11), 3141-3148; (b) Muratov, C. B.; Shvartsman, S. Y., Boundary homogenization for periodic arrays of absorbers. *Multiscale Model Sim* **2008,** *7* (1), 44-61; (c) Martin, P. A.; Skvortsov, A. T., Scattering by a sphere in a tube, and related problems. *J Acoust Soc Am* **2020,** *148* (1), 191-200; (d) Martin, P. A.; Skvortsov, A. T., On blockage coefficients: flow past a body in a pipe. *Proc. R. Soc. A* **2022,** *478*, 2021.0677.